\documentclass{blois}
\usepackage{amsmath}
\newcommand{\Z}{\mathrm{Z}}
\newcommand{\W}{\mathrm{W}}

\def\be{\begin{equation}}
\def\ee{\end{equation}}
\def\bea{\begin{eqnarray}}
\def\eea{\end{eqnarray}}

\begin{document}
\vspace*{4cm}
\title{Higgs differential cross section and STXS measurements at CMS}

\author{ Tahir Javaid~\footnote{\href{mailto:tahir.javaid@cern.ch}{tahir.javaid@cern.ch}} , Li Yuan~\footnote{\href{mailto:li.yuan@cern.ch}{li.yuan@cern.ch}} and Tongguang Cheng~\footnote{\href{mailto:tongguang.cheng@cern.ch}{tongguang.cheng@cern.ch}} \\
Beihang University, China}

\address{On the behalf of CMS Collaboration of CERN,\\
Geneva-Switzerland}

\def\Photo{}

\maketitle 
\abstracts{
In this manuscript, we present the latest differential measurements of Higgs boson cross sections with the CMS detector in bosonic and fermionic decay channels. Both fiducial differential cross section measurements and measurements in the simplified template cross section framework are presented. The fiducial measurements are then used to compute limits on Higgs couplings using the Standard Model Effective Field Theory. The results are based on data collected during Run 2 of the LHC by the CMS experiment. First set of differential measurements with early Run 3 data are also reported. 
}

\section{Introduction}

The discovery of the Higgs boson by the ATLAS~\cite{ATLAS:2012yve} and CMS~\cite{CMS:2012qbp}  experiments marked a crucial milestone in precision physics, thus enabling detailed studies of its properties including its production cross section. To have deeper insights into the Standard Model (SM), fiducial and differential cross section measurements play a pivotal role. Fiducial measurements offer a way to determine production cross-sections independently of theoretical models while also improving sensitivity to possible beyond-the-Standard Model (BSM) effects. Such measurements confine extrapolations to a precisely defined phase space that closely match the experimental selections, thereby minimizing reliance on theoretical assumptions. This facilitates direct comparisons with different theoretical predictions. On the other hand, differential measurements divide the cross-section into bins based on specific kinematic observables, yielding more granular insights than inclusive measurements.
An alternative but complementary strategy is the simplified template cross-sections (STXS) framework, which categorizes cross-sections into predefined production bins. This method enhances experimental sensitivity to potential BSM phenomena while reducing dependence on theoretical models. 

This document provides an overview of recent differential and STXS cross-section measurements of Higgs boson production, conducted by the CMS Collaboration. The differential measurements are reported in Higgs bosonic and fermionic decay channels and as well as from the combination of several decay channels. The combined spectra provide measurements at the highest level of precision presently achievable. The results are derived from data collected during LHC Run 2 at $\sqrt{s} = 13$~TeV, corresponding to an integrated luminosity of $\mathcal{L} = 138$~fb$^{-1}$. Additionally, preliminary differential measurements from early Run 3 CMS data, performed at a slightly higher energy of $\sqrt{s} = 13.6$~TeV with an integrated luminosity of $\mathcal{L} = 34.7$~fb$^{-1}$, are also presented. A first set of studies on interpretations of the combined differential spectra in terms of Higgs boson couplings using the Standard Model Effective Field Theory (SMEFT) approach are provided. %The measurement is done by fitting CP-even CP-odd pairs of coefficients using transverse momentum of Higgs ($p_{T}^{H}$).

\section{Higgs Cross section measurement in bosonic final states }
In this section, recent results of $\mathrm{H}\rightarrow{\gamma \gamma}$ \cite{CMS:2024pfc} and $\mathrm{H}\rightarrow{\rm Z}{\rm Z}(\rightarrow4\ell$ ($\ell={\rm e},\mu$)) \cite{CMS:2025wnr} decay channels are described. The results are obtained using early Run 3 data recorded at CMS detector of LHC at 13.6 TeV. Despite some improvements in the analyses’ components, the overall strategy of the measurement for these analyses is similar to their counterpart using the Run 2 CMS data.  
\subsection{Differential measurements in $\mathrm{H}\rightarrow{\gamma \gamma}$ decay channel}
\begin{figure}
\begin{minipage}{0.325\linewidth}
\centerline{\includegraphics[width=0.97\linewidth]{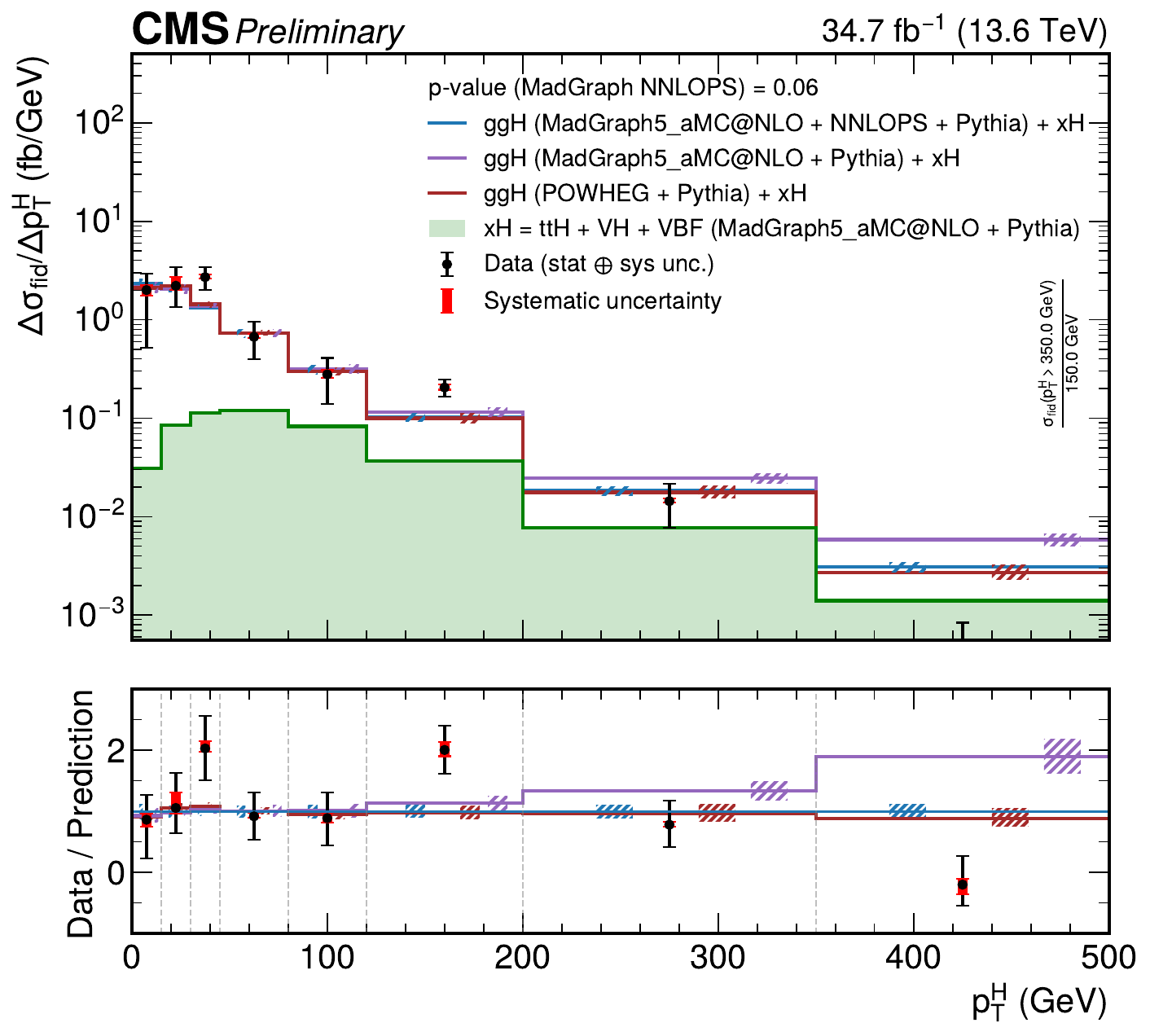}}
\end{minipage}
\hfill
\begin{minipage}{0.325\linewidth}
\centerline{\includegraphics[width=0.97\linewidth]{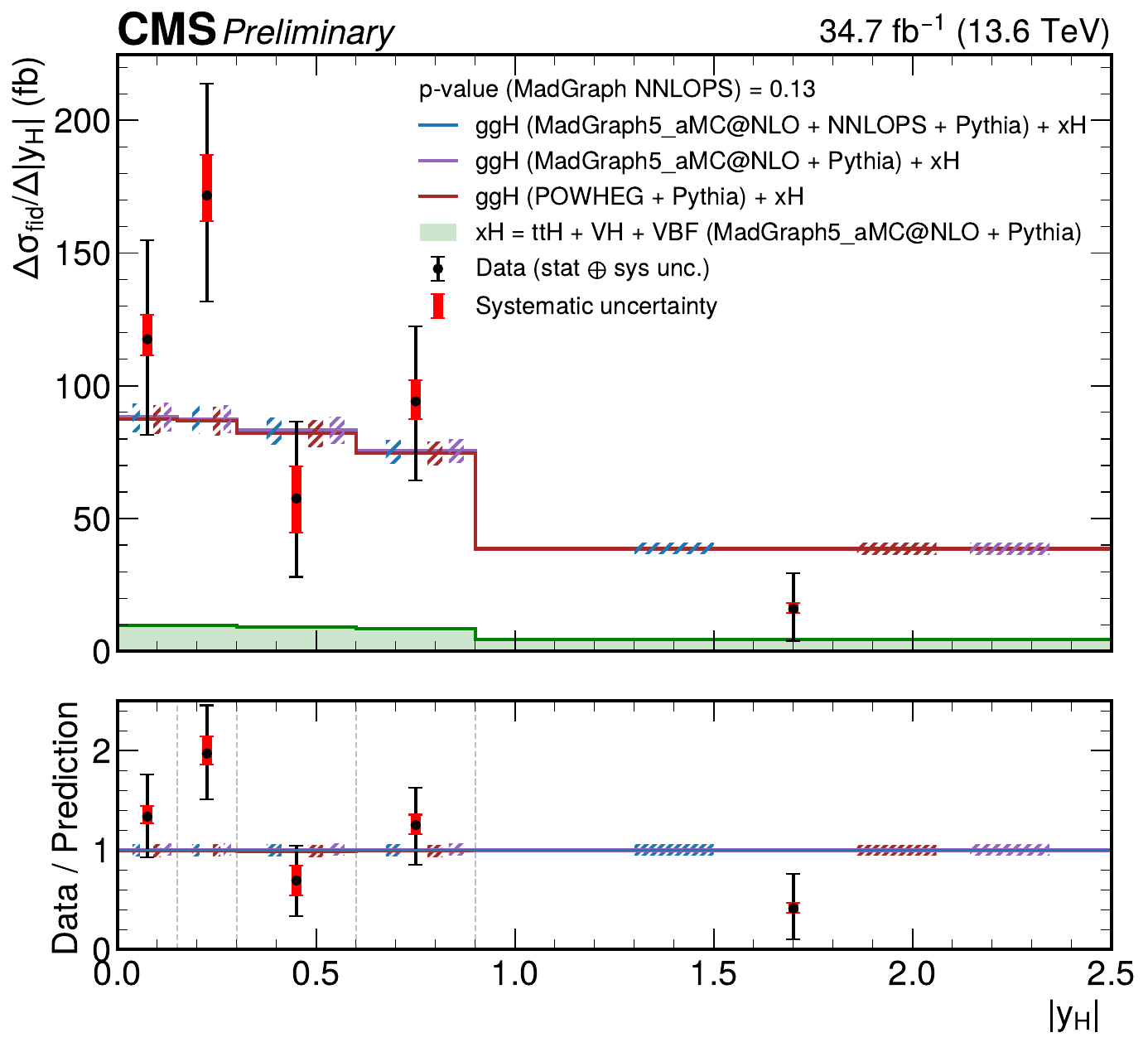}}
\end{minipage}
\hfill
\begin{minipage}{0.325\linewidth}
\centerline{\includegraphics[width=0.97\linewidth]{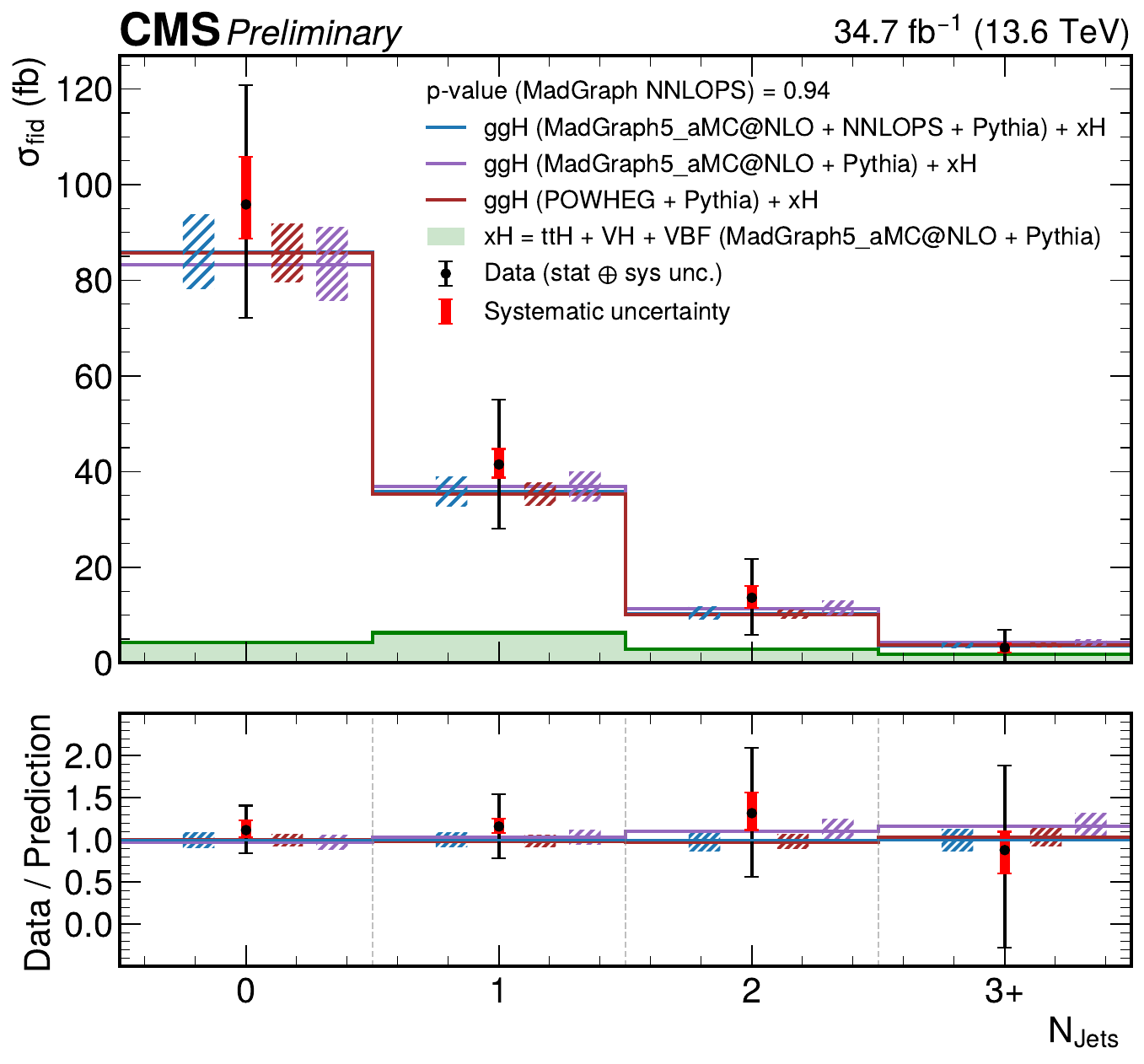}}
\end{minipage}
\caption[]{The differential fiducial cross-sections for $p_{\text{T}}^{\text{H}}$ (left), $|y^H|$ (center) and $N_{Jets}$ (right) in $\mathrm{H}\rightarrow{\gamma \gamma}$ decay channel are presented. 
%Various event generation setups, detailed in the legend and text, are represented by different colored lines. Dashed boxes illustrate the uncertainties in theoretical predictions for both the ggH and xH components. The provided $p$-value corresponds to the nominal Standard Model (SM) prediction. The bottom panel displays the ratio relative to the nominal SM prediction. In the $p_{\text{T}}^{\text{H}}$ distribution, the final bin extends to infinity, with its normalization specified in the plot.
}
\label{fig:higgs_gamma_differential}
\end{figure}

In contrast to $\mathrm{H}\rightarrow{\rm Z}{\rm Z}(\rightarrow4\ell$ ($\ell={\rm e},\mu$)), $\mathrm{H}\rightarrow{\gamma \gamma}$ channel has lower sensitivity however it offers excellent data-driven background estimation. For the measurement, there are certain requirements on the objects (i.e. transverse momentum of photons, supercluster pseudo-rapidity, photon identification, shower shape and isolation variables).  With respect to previous measurement, the  fiducial requirements are updated for improved perturbative convergence in phase space. 
To measure the fiducial cross-section, $\sigma_{\text{fid}}$, a profile-likelihood fit is applied to the diphoton invariant mass ($m_{\gamma\gamma}$) distributions. This fit is performed within three categories based on mass-resolution. The results are reported for three observables related to Higgs production as shown in Figure. \ref{fig:higgs_gamma_differential}. The dominant systematic uncertainty for this measurement is photon energy scale and resolution. 

\subsection{Differential measurements in $\mathrm{H}\rightarrow{\rm Z}{\rm Z}$ decay channel}
\begin{figure}
\begin{minipage}{0.49\linewidth}
\centerline{\includegraphics[width=0.775\linewidth]{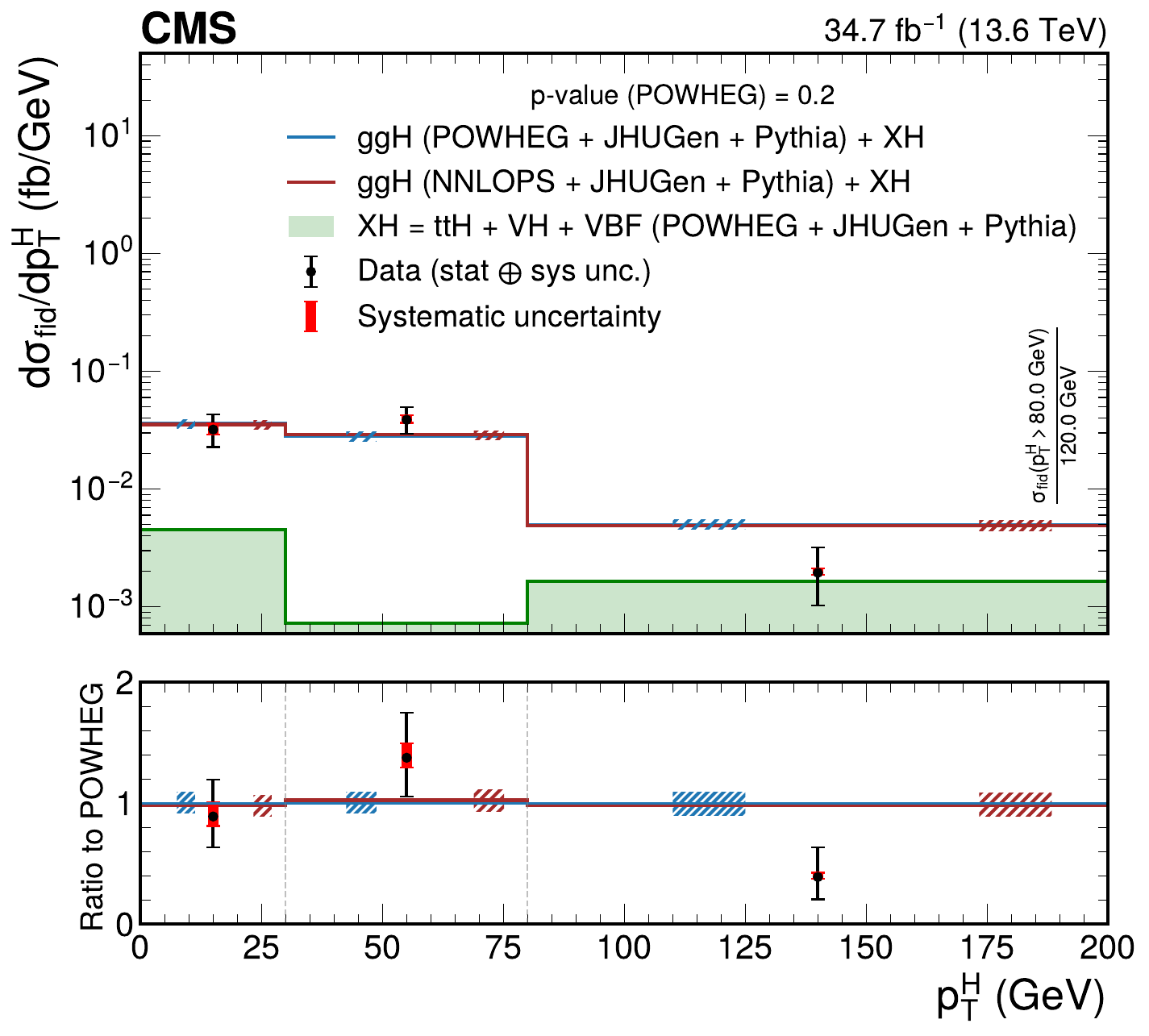}}
\end{minipage}
\hfill
\begin{minipage}{0.49\linewidth}
\centerline{\includegraphics[width=0.775\linewidth]{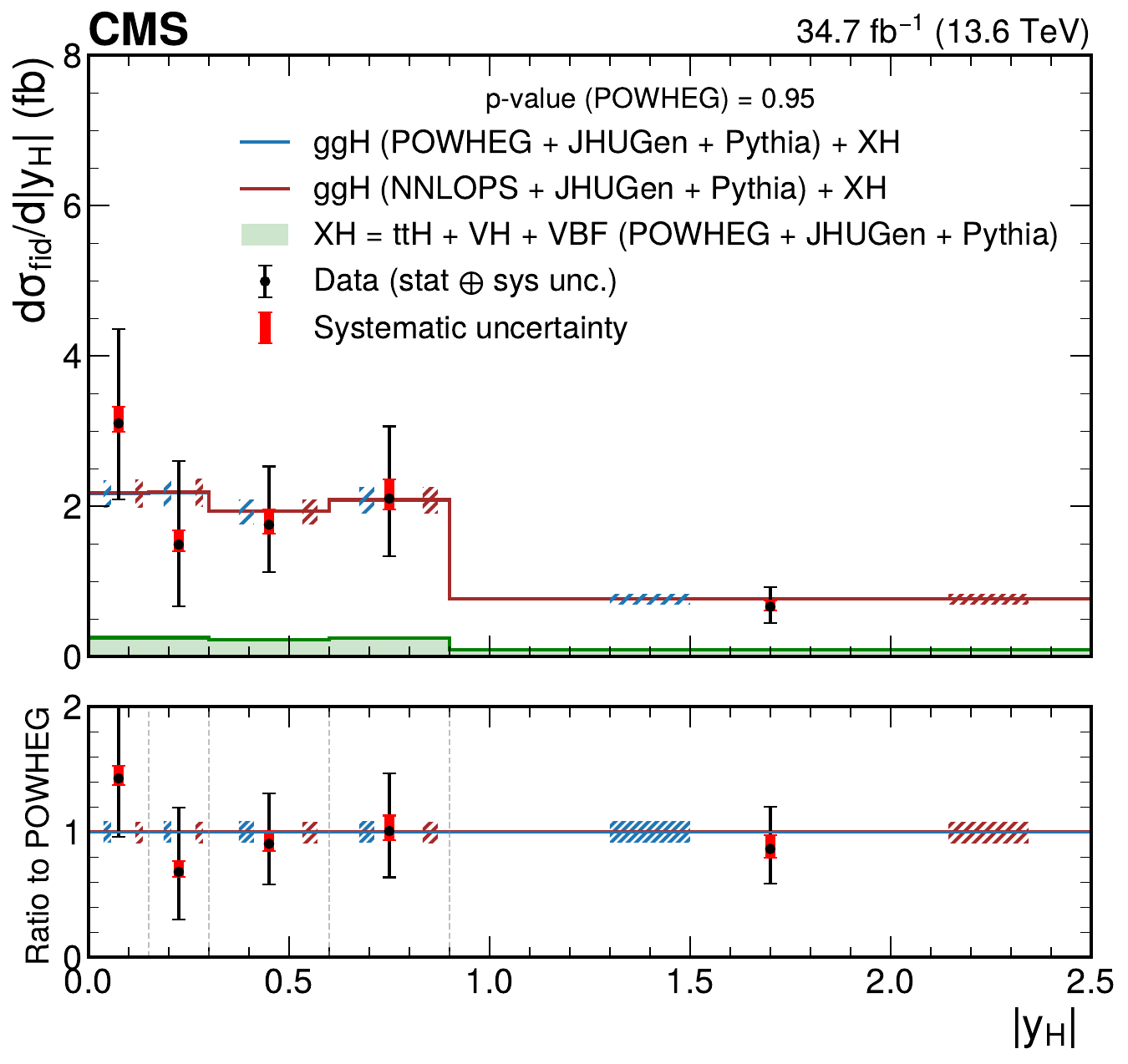}}
\end{minipage}
%\begin{minipage}{0.32\linewidth}
%\centerline{\includegraphics[angle=-45,width=0.7\linewidth]{figexamp}}
%\end{minipage}
\caption[]{The differential fiducial cross-sections for $p_{\text{T}}^{\text{H}}$ (left) and $|y^H|$ (right) in $\mathrm{H}\rightarrow{\rm Z}{\rm Z}$ decay channel are presented. 
%Various event generation setups, detailed in the legend and text, are represented by different colored lines. Dashed boxes illustrate the uncertainties in theoretical predictions for both the ggH and xH components. The provided $p$-value corresponds to the nominal Standard Model (SM) prediction. The bottom panel displays the ratio relative to the nominal SM prediction. In the $p_{\text{T}}^{\text{H}}$ distribution, the final bin extends to infinity, with its normalization specified in the plot.
}
\label{fig:higgs_zz_differential}
\end{figure}

This channel is much suited for differential measurement as it offers rather clean signal. 
%Overall strategy for background estimation and result extraction is similar to what has already been reported in recent studies with Run2 dataset \cite{CMS:2023gjz}. 
The analysis follows several requirements on lepton kinematics (i.e. $p_T, |\eta|, ID $) and isolation to match the HLT requirements. The differential fiducial cross-section for the process $pp \to H \to 4\ell$ is extracted using an unbinned maximum likelihood fit performed over observed four lepton invariant mass in the final state. This fit accounts for both the signal and background contributions in the observed $4\ell$ mass distribution, $N_{\text{obs}}(m_{4\ell})$, from which the fiducial cross-section $\sigma_{\text{fid}}$ is directly extracted in bins of $p_{\text{T}}^{\text{H}}$ and $|y^H|$ observables. %The Higgs boson mass, $m_H$, is fixed at 125.38 GeV in the fit. 
The measured cross section is shown in Figure. \ref{fig:higgs_zz_differential}.

\section{Higgs cross section measurement in fermionic final states }

In fermionic final states, measurements are reported for $\mathrm{H}\rightarrow{\rm b}{\rm b}$ and $\mathrm{H}\rightarrow{\rm \tau}{\rm \tau}$ (having SM branching ratios of nearly 58\% and 6\% respectively) decay channels. The $\mathrm{H}\rightarrow{\rm b}{\rm \bar{b}}$ channel dominates but is challenging to observe due to large QCD backgrounds, requiring advanced tagging techniques. The $\mathrm{H}\rightarrow{\rm \tau}{\rm \tau}$ channel benefits from a cleaner final state with distinct kinematic features, enabling more precise measurements. These two channels provide crucial tests of Higgs couplings to fermions, complementing bosonic decay modes. 

\subsection{Cross sections measurements in $\mathrm{H}\rightarrow{\rm b}{\rm \bar{b}}$ decay channel}
\begin{figure}
\begin{minipage}{0.49\linewidth}
\centerline{\includegraphics[width=0.82\linewidth]{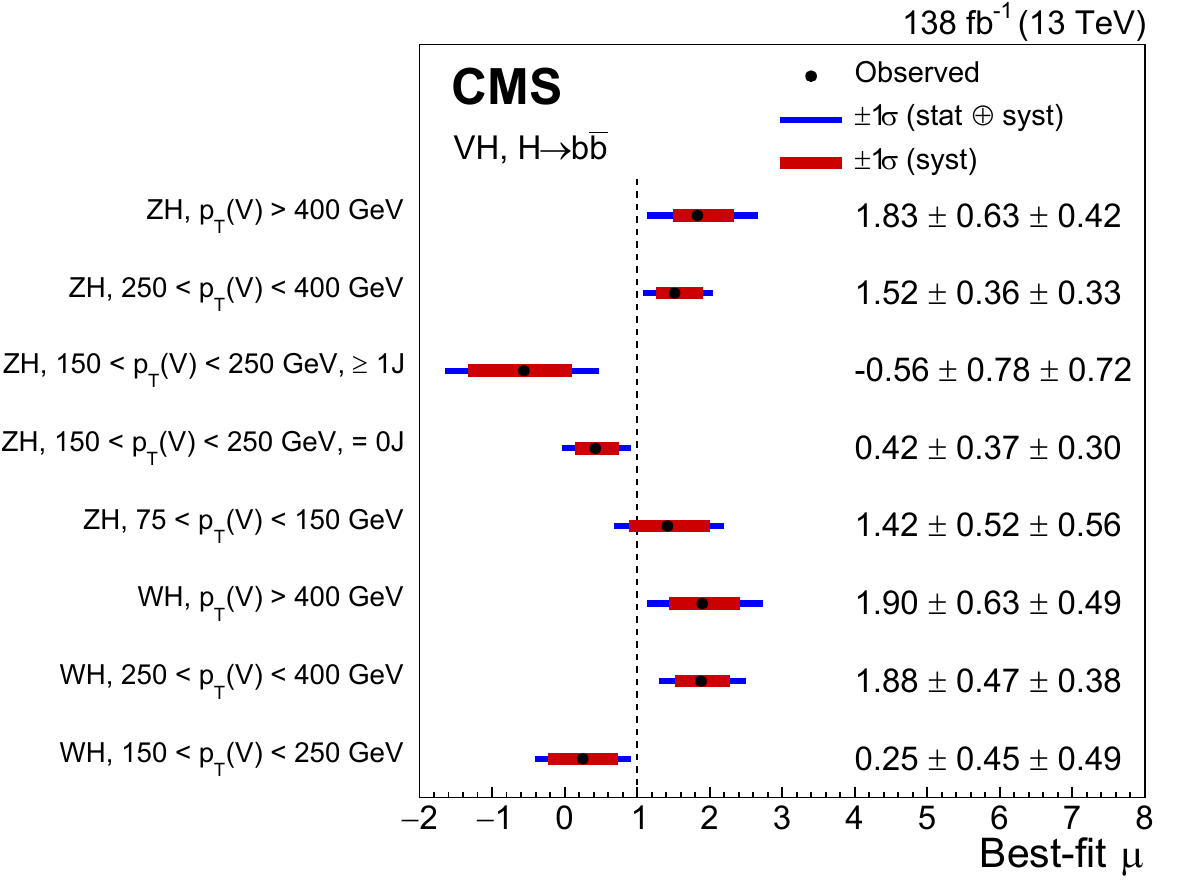}}
\end{minipage}
\hfill
\begin{minipage}{0.49\linewidth}
\centerline{\includegraphics[width=0.82\linewidth]{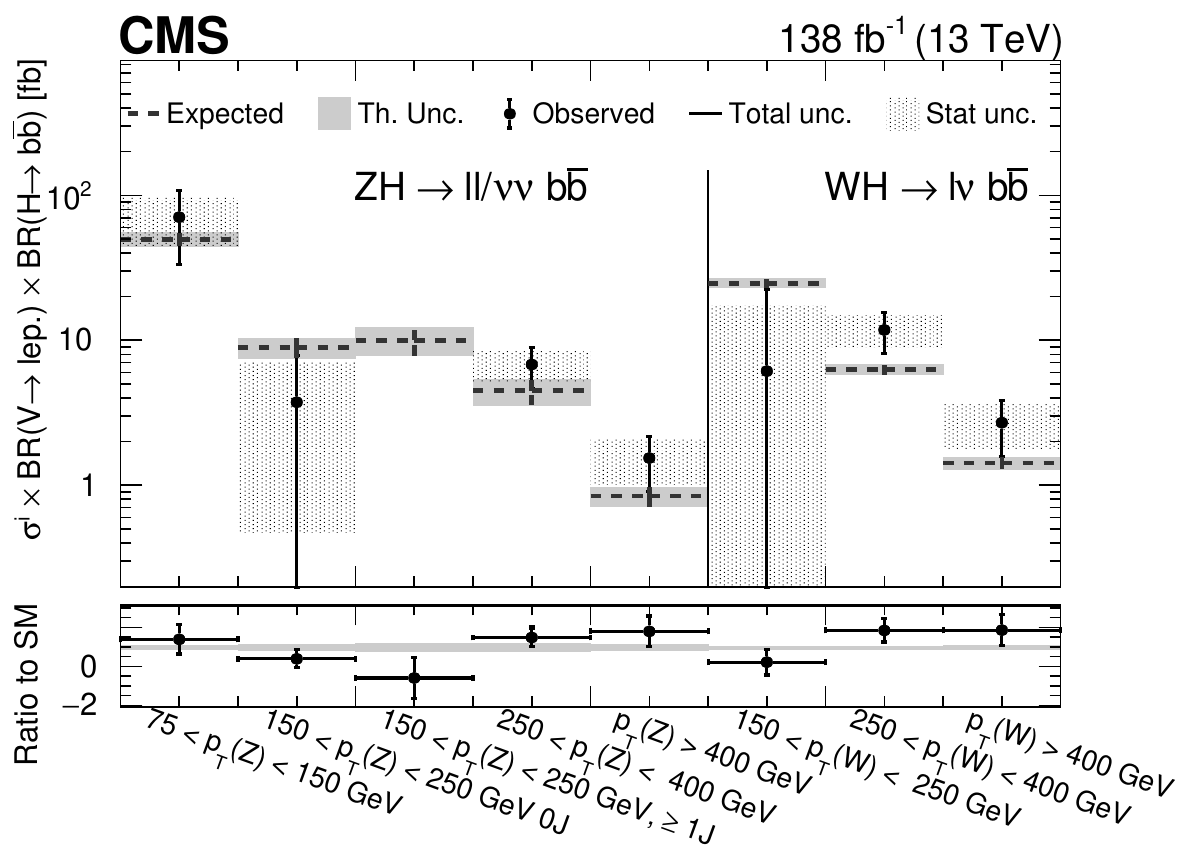}}
\end{minipage}
\hfill
\caption[]{The measured STXS signal strengths are presented (left). A vertical dashed line represents the SM expectation for the signal strength. 
%The first and second uncertainty values correspond to statistical and systematic uncertainties, respectively. 
The measured values of $\sigma \times \mathcal{B}$ are also reported (right).
%%, where $\sigma \times \mathcal{B}$ is defined as the product of the $VH$ production cross-sections and the branching fractions for $V \to$ leptons and $H \to b\bar{b}$. .
%%These measurements are evaluated using the same STXS bins as the signal strengths, with all years combined.
%The lower panel presents the ratio of the observed results, including associated uncertainties, to the SM expectations. 
%%If the observed signal strength in a particular STXS bin is negative, no value is plotted for $\sigma \mathcal{B}$ in the upper panel.
}
\label{fig:Hbb_VH}
\end{figure}

In this section, several measurements on Higgs cross section are reported. Firstly, differential cross-sections are evaluated for SM Higgs boson production in association with vector bosons ($\W$ or~ $\Z$), where the Higgs decays into a pair of $b$ quarks \cite{CMS:2023vzh}. These measurements are conducted following the simplified template cross-section framework and the analysis utilizes the leptonic decays of the $W$ and $Z$ bosons, leading to final states containing 0, 1, or 2 electrons or muons.  Higgs boson candidates are reconstructed either from resolved pairs of $b$-tagged jets or from single large-radius jets that encapsulate particles originating from two $b$ quarks. Cross section is extracted by performing fit to signal region and orthogonal control regions (CRs) defined with the help of Deep Neural Network (DNN). The results are shown in the Figure. \ref{fig:Hbb_VH}. Simulation Modeling, b-tagging, JER being leading systematic sources for this measurement.

\begin{figure}
\begin{minipage}{0.49\linewidth}
\centerline{\includegraphics[width=0.8\linewidth]{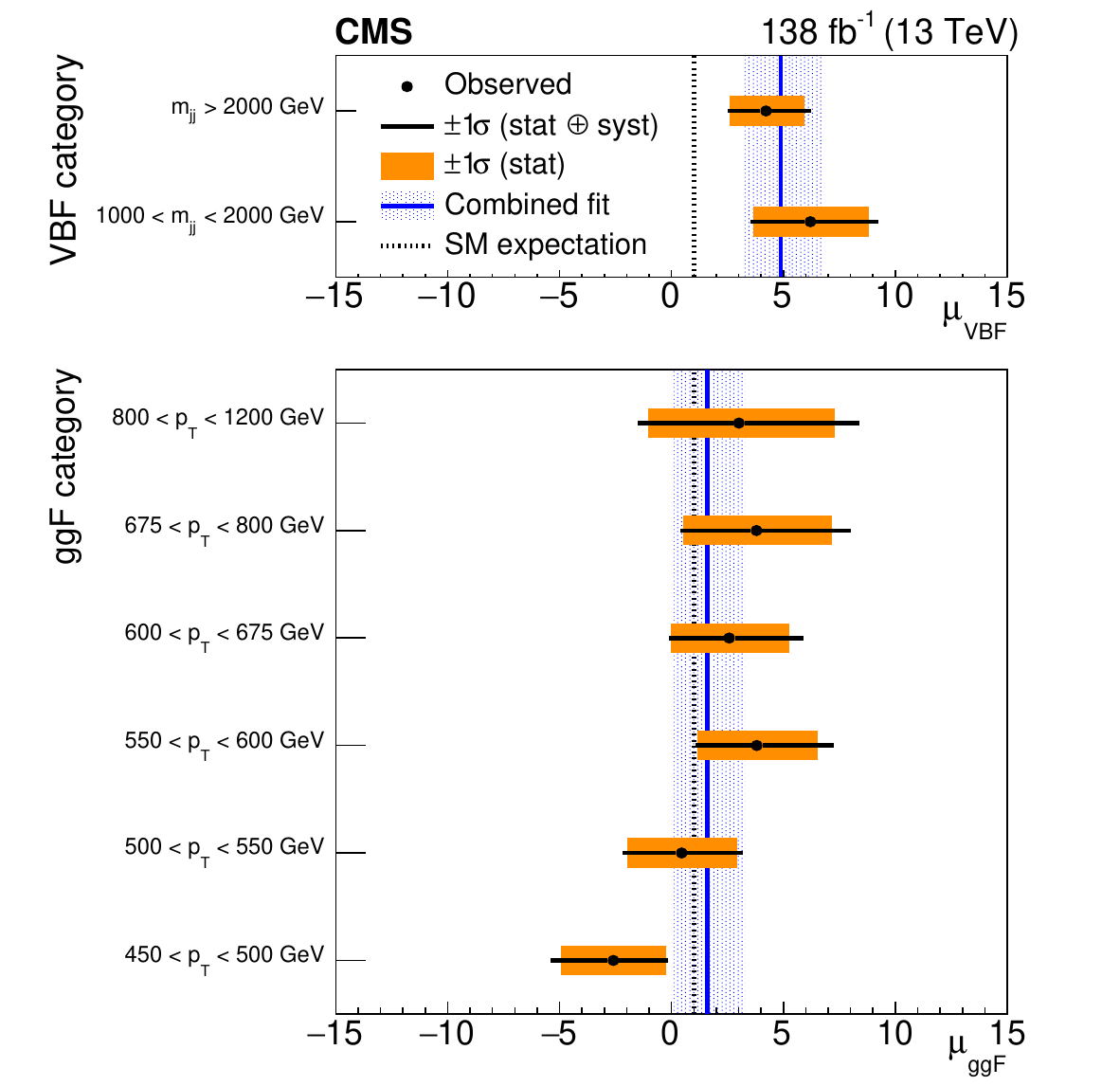}}
\end{minipage}
\hfill
\begin{minipage}{0.49\linewidth}
\centerline{\includegraphics[width=0.85\linewidth]{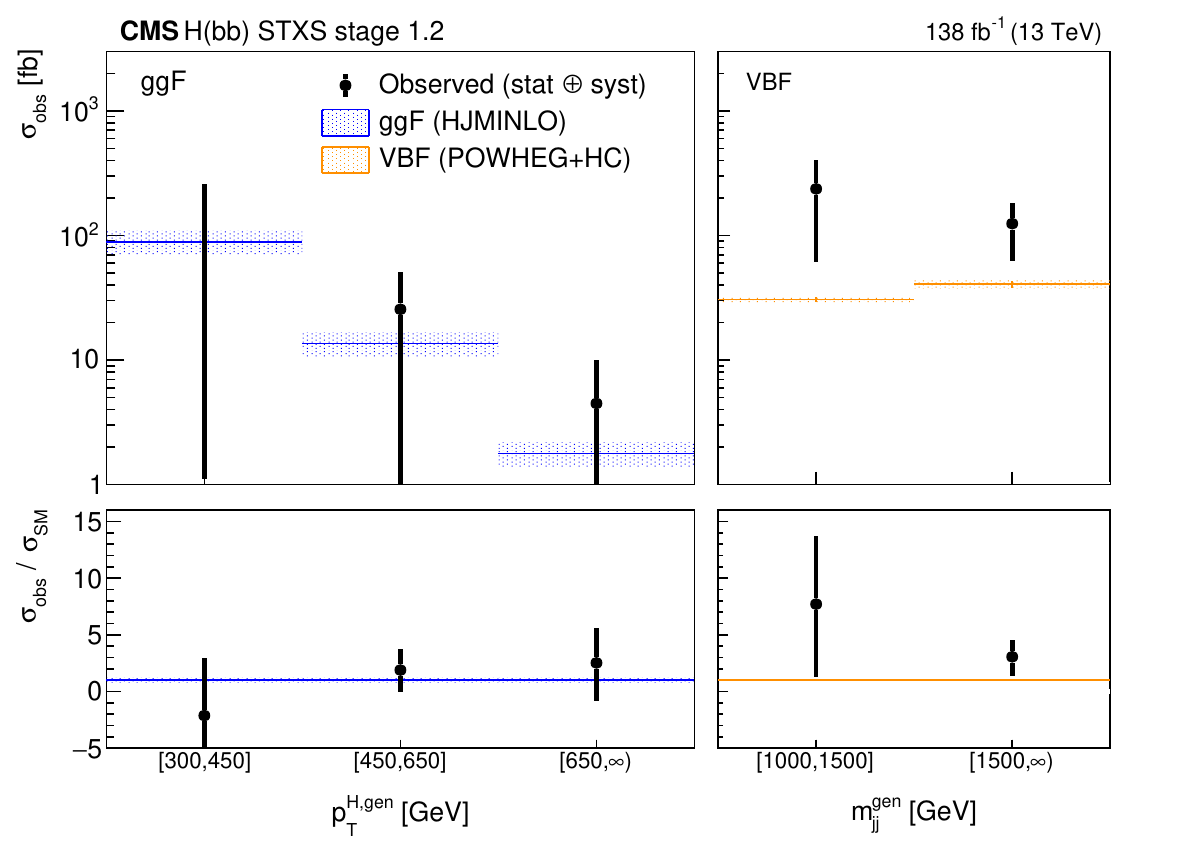}}
\end{minipage}
\hfill
\caption[]{VBF and ggF signal strengths measured in bins of $m_{jj}$ and $p_T$ are shown (left). The unfolded STXS is also presented (right) for three bins of Higgs boson $p_T$ in ggF and two bins of generator-level $m_{jj}$ in VBF. 
%%The Standard Model (SM) prediction from HJMINLO and POWHEG, incorporating higher-order electroweak (EW) and next-to-next-to-leading order (NNLO) QCD corrections, is overlaid for the ggF and VBF production mechanisms. 
%The lower panel illustrates the ratio of the measured cross-section to the SM expectation.
}
\label{fig:Hbb_boosted}
\end{figure}

Further, boosted $H \to b\bar{b}$ decays \cite{CMS:2024ddc}, marking the first investigation of $H \to b\bar{b}$ in the VBF or ggF production mode is reported. This study employs an advanced version of the tagger algorithm for identifying $H \to b\bar{b}$ decays and rejecting the major backgrounds. Key backgrounds for this analysis include QCD multijet production and top quark pair ($t\bar{t}$) production. The latter is normalized through a dedicated muon control region, incorporated into the final likelihood fit of the invariant $H \to b\bar{b}$ mass. Measured cross sections are reported in Firgure. \ref{fig:Hbb_boosted}.

\begin{figure}
\begin{minipage}{0.325\linewidth}
\centerline{\includegraphics[width=0.95\linewidth]{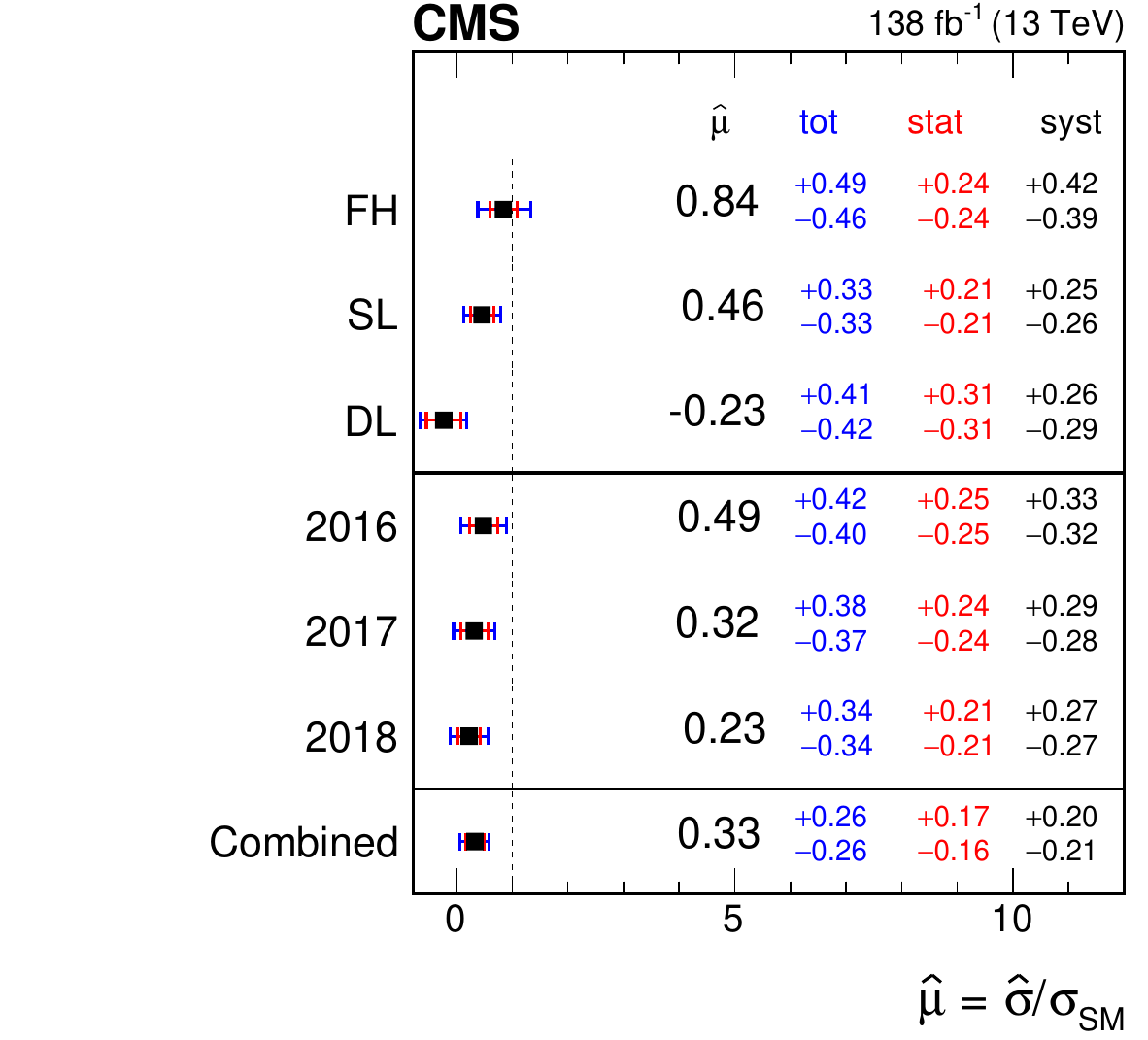}}
\end{minipage}
\hfill
\begin{minipage}{0.325\linewidth}
\centerline{\includegraphics[width=0.95\linewidth]{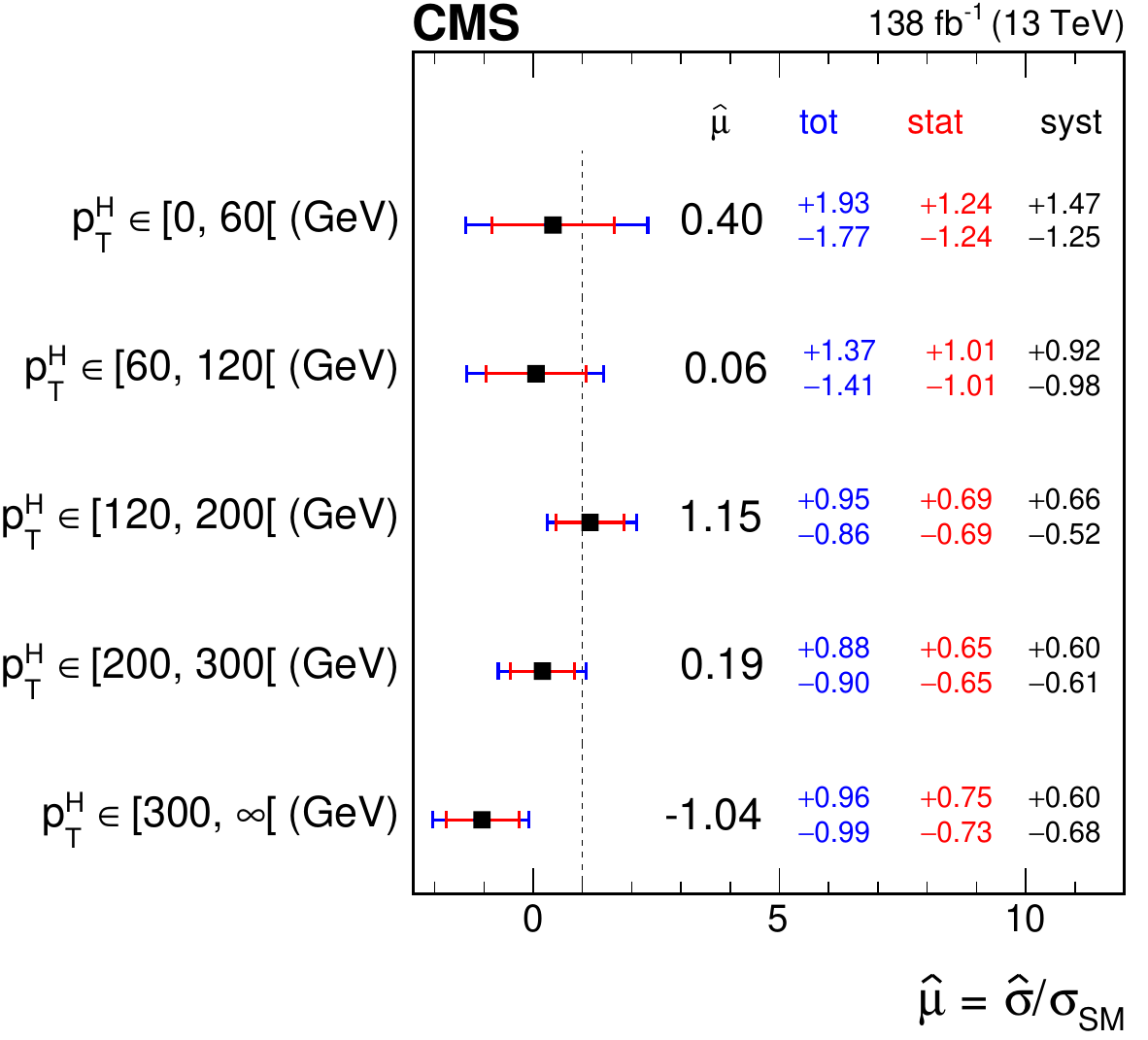}}
\end{minipage}
\hfill
\begin{minipage}{0.325\linewidth}
\centerline{\includegraphics[width=0.95\linewidth]{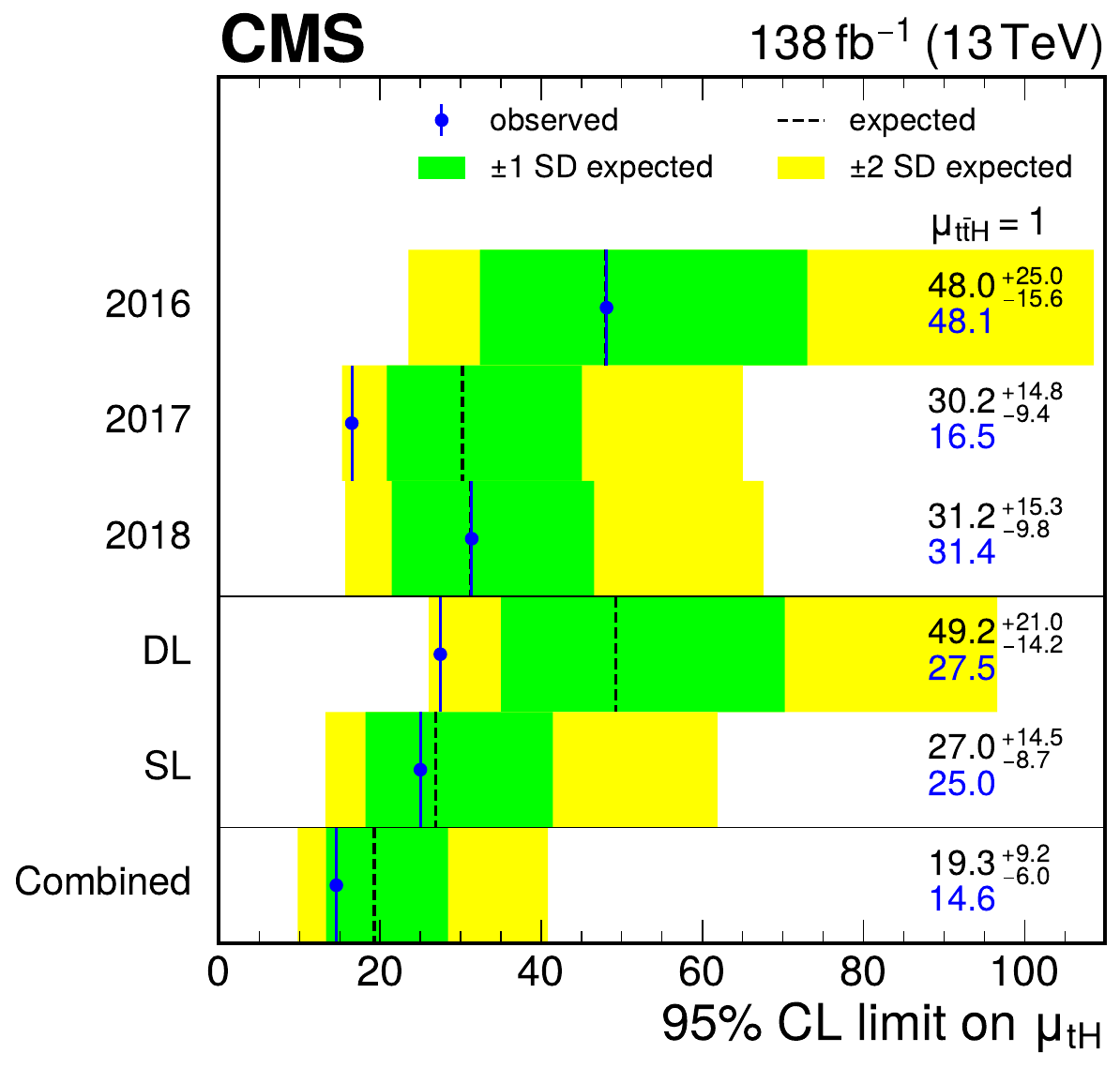}}
\end{minipage}
\caption[]{The $t\bar{t}H$ signal strength $\mu_{t\bar{t}H}$ are shown inclusively (left) and in different $p_{\text{T}}^{H}$ bins from the STXS measurement (center). 
% for each analysis channel (top three rows), for each data-taking year (middle three rows), and for the combined result across all channels and years (bottom row). 
%The uncertainties are correlated across different channels and years. 
%The best-fit values of the $t\bar{t}H$ signal strength modifiers $\mu_{t\bar{t}H}$ are shown (center) for different $p_{\text{T}}^{H}$ bins (left)
%as obtained from the STXS measurement
The upper 95\% confidence level limit on the $tH$ signal strength is also presented (right)
%for various channels and years
.
%The observed limit is shown as a solid vertical line, while the expected limit is depicted as a dashed vertical line. 
%In this measurement, uncertainties are treated as uncorrelated across different channels and years, as well as in their combined result. The green and yellow shaded regions represent the one- and two-standard deviation confidence intervals on the expected limit, respectively.
}
\label{fig:Hbb_tH_ttH}
\end{figure}

Another measurement is reported where Higgs is produced in association with single top quark ($tH$) or a pair of top quarks ($t\bar{t}H$) \cite{CMS:2023vtj} where are all possible decay channels for the $t\bar{t}$ system are studied. Selected events are categorized on the basis of jet and b-tag multiplicity in which artificial neural network (ANN) output is used to separate the signal and background. Major background for this analysis comes from $t\bar{t}+jets$ QCD process which is estimated from data. The production rates of the $t\bar{t}H$ and $tH$ signal processes are obtained using a simultaneous binned profile likelihood fit applied to the final discriminant distributions. This fit incorporates all channels (i.e. Fully Hadronic (FH), Semi leptinic (SL) and Dileptonic (DL) ), event categories, and data-taking periods, following the methodologies described in Ref. \cite{CMS:2023vtj}. The results are shown in the Figure. \ref{fig:Hbb_tH_ttH}.

\subsection{Cross sections measurements in $\mathrm{H}\rightarrow{\rm \tau}{\rm \bar{\tau}}$ decay channel}
\begin{figure}
\begin{minipage}{0.325\linewidth}
\centerline{\includegraphics[width=0.97\linewidth]{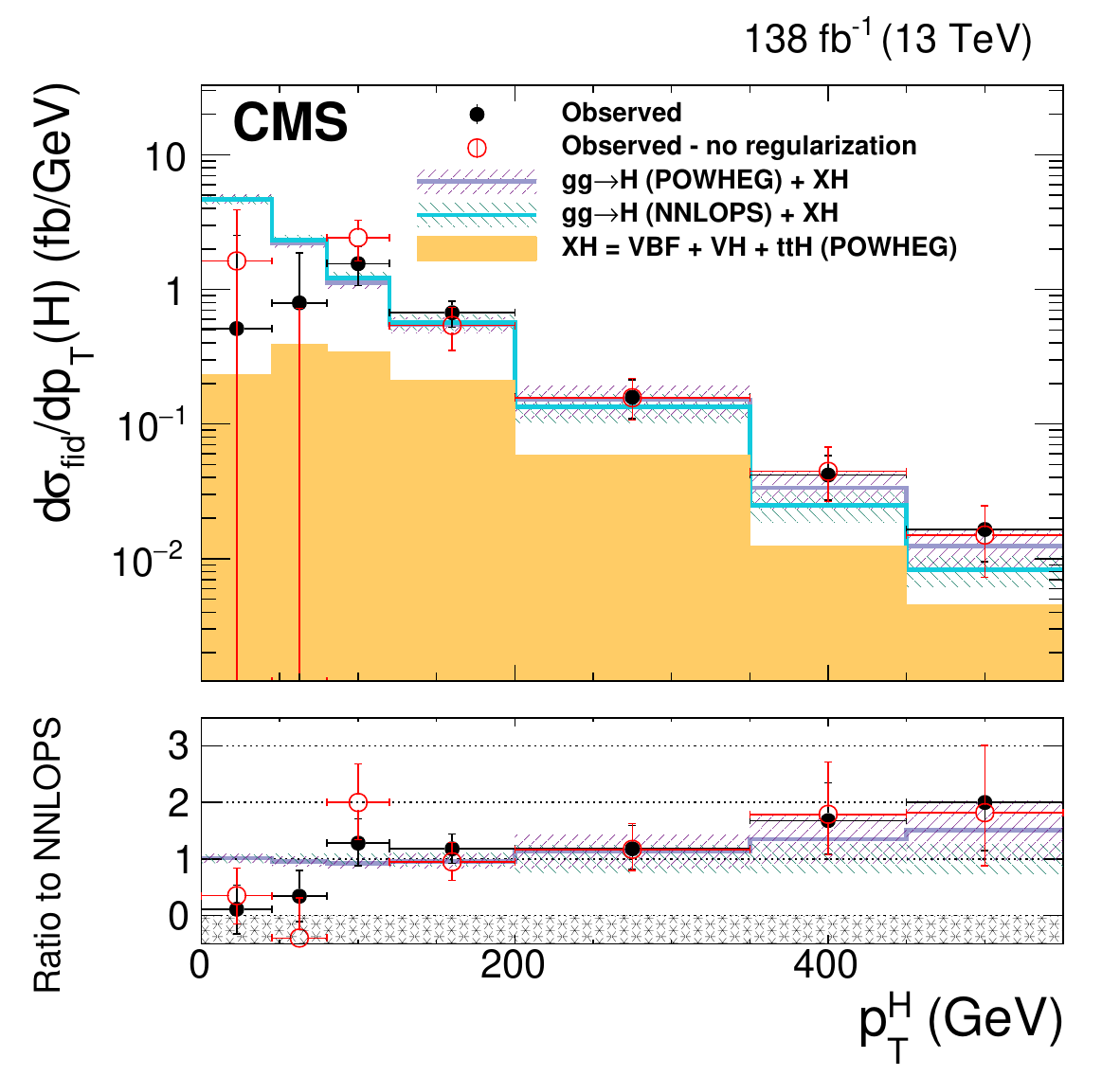}}
\end{minipage}
\hfill
\begin{minipage}{0.325\linewidth}
\centerline{\includegraphics[width=0.97\linewidth]{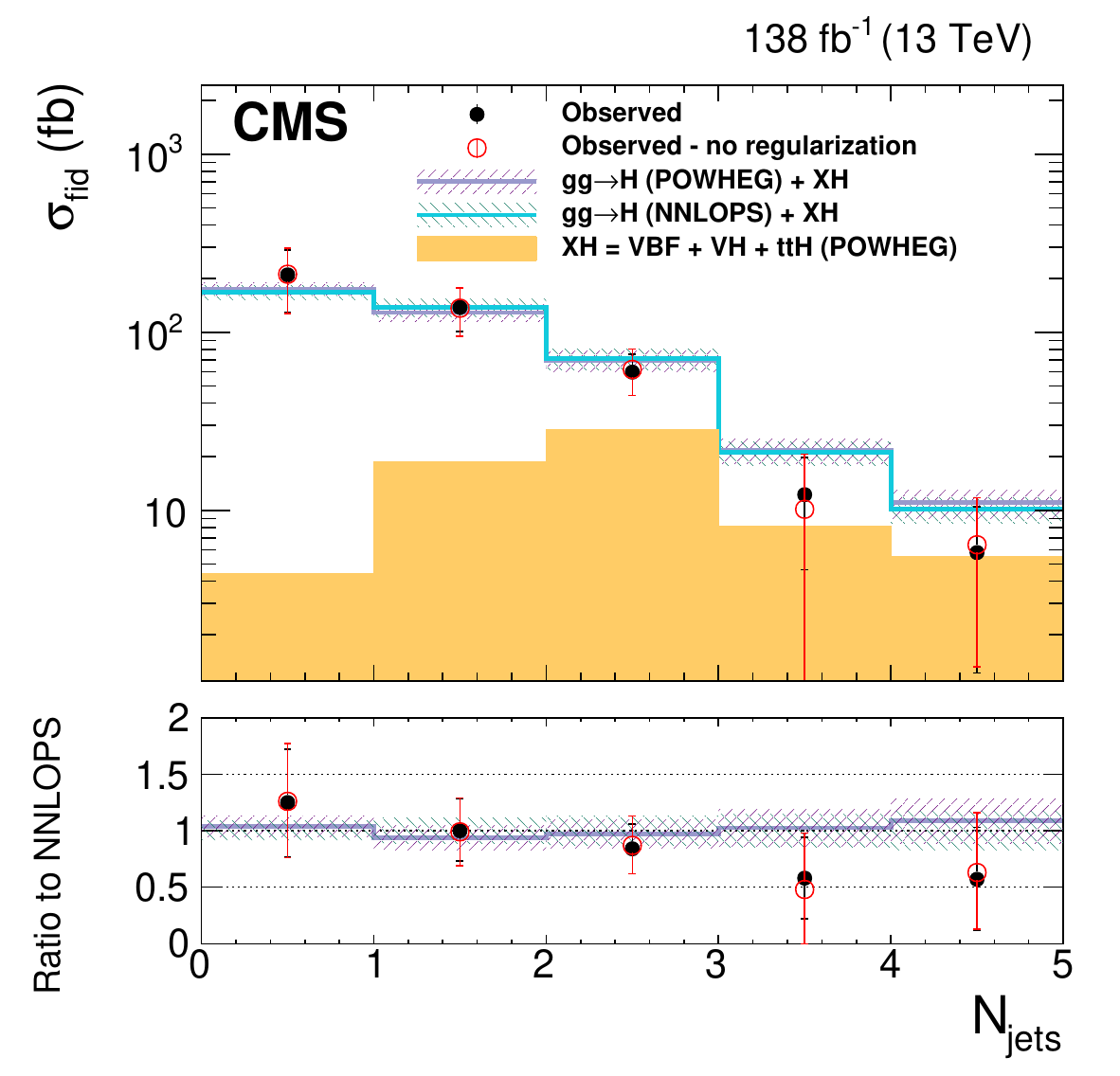}}
\end{minipage}
\hfill
\begin{minipage}{0.325\linewidth}
\centerline{\includegraphics[width=0.97\linewidth]{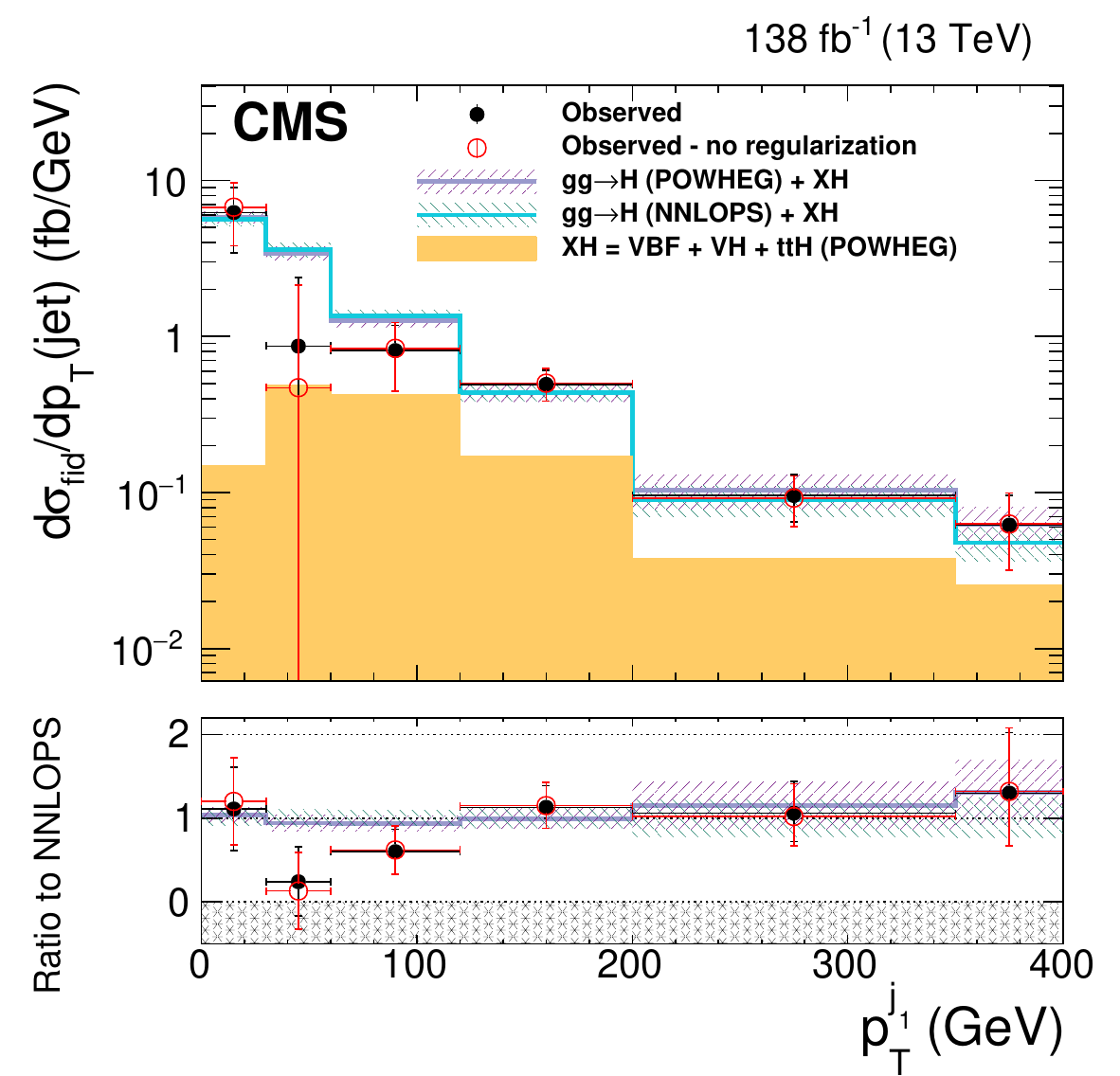}}
\end{minipage}
\caption[]{The observed and expected differential fiducial cross-sections are shown in bins of $p_{\text{T}}^{H}$ (left), $N_{\text{jets}}$ (center), and $p_{\text{T}}^{j_1}$ (right). 
%Both regularized (full markers) and unregularized (hollow markers) data points are included. The leftmost bin in the $p_{\text{T}}^{j_1}$ distribution accounts for events without a jet with $p_T > 30$ GeV.
%The uncertainty bands in the theoretical predictions incorporate contributions from various sources, including the parton distribution function (PDF), renormalization and factorization scales, underlying event modeling, parton showering, and the Higgs boson branching fraction to $\tau$ leptons. 
The last bins also include the overflow.}
\label{fig:Htautau_differential_resolved}
\end{figure}

\begin{figure}
\begin{minipage}{0.49\linewidth}
\centerline{\includegraphics[width=0.8\linewidth]{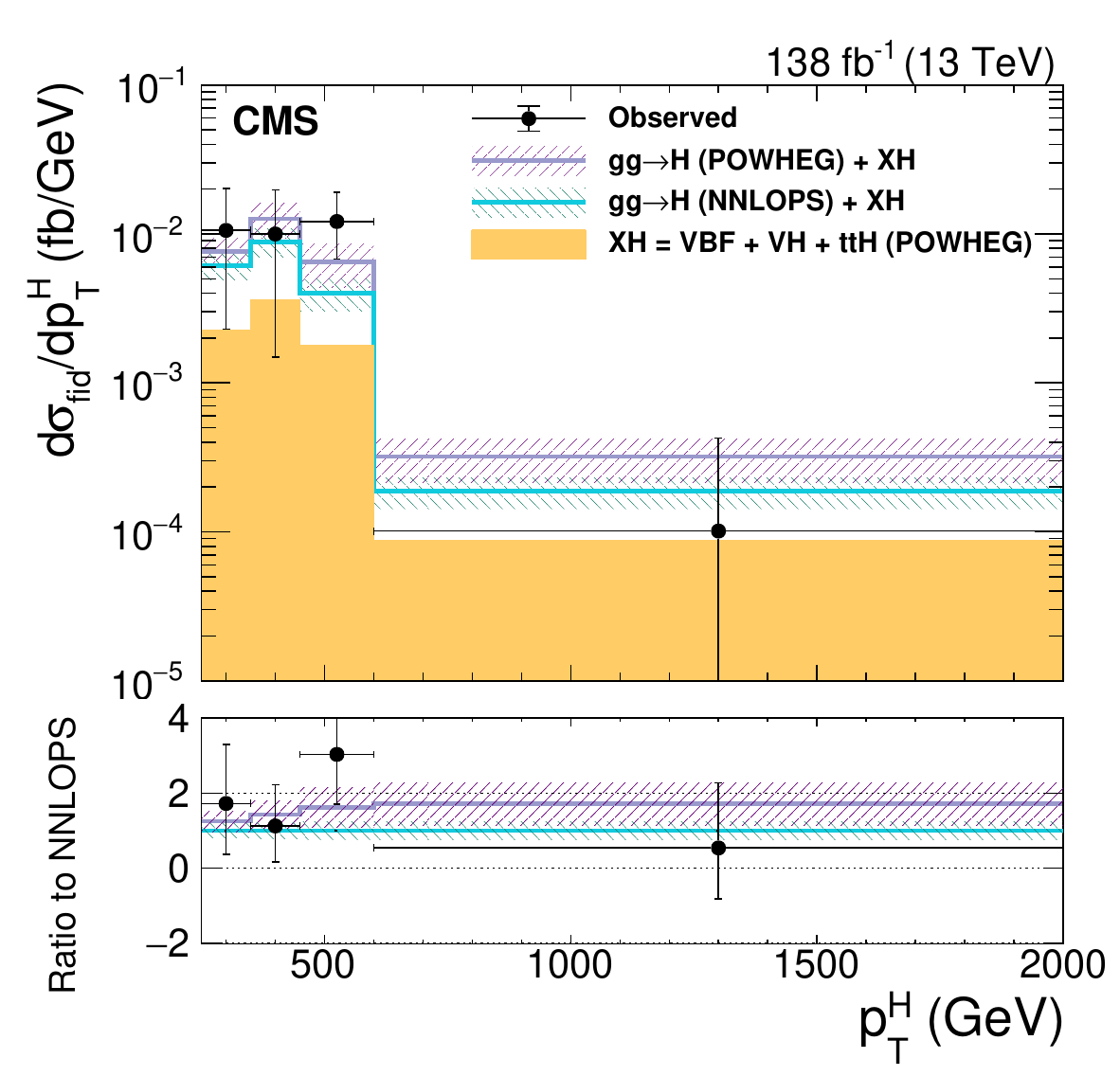}}
\end{minipage}
\hfill
\begin{minipage}{0.49\linewidth}
\centerline{\includegraphics[width=0.8\linewidth]{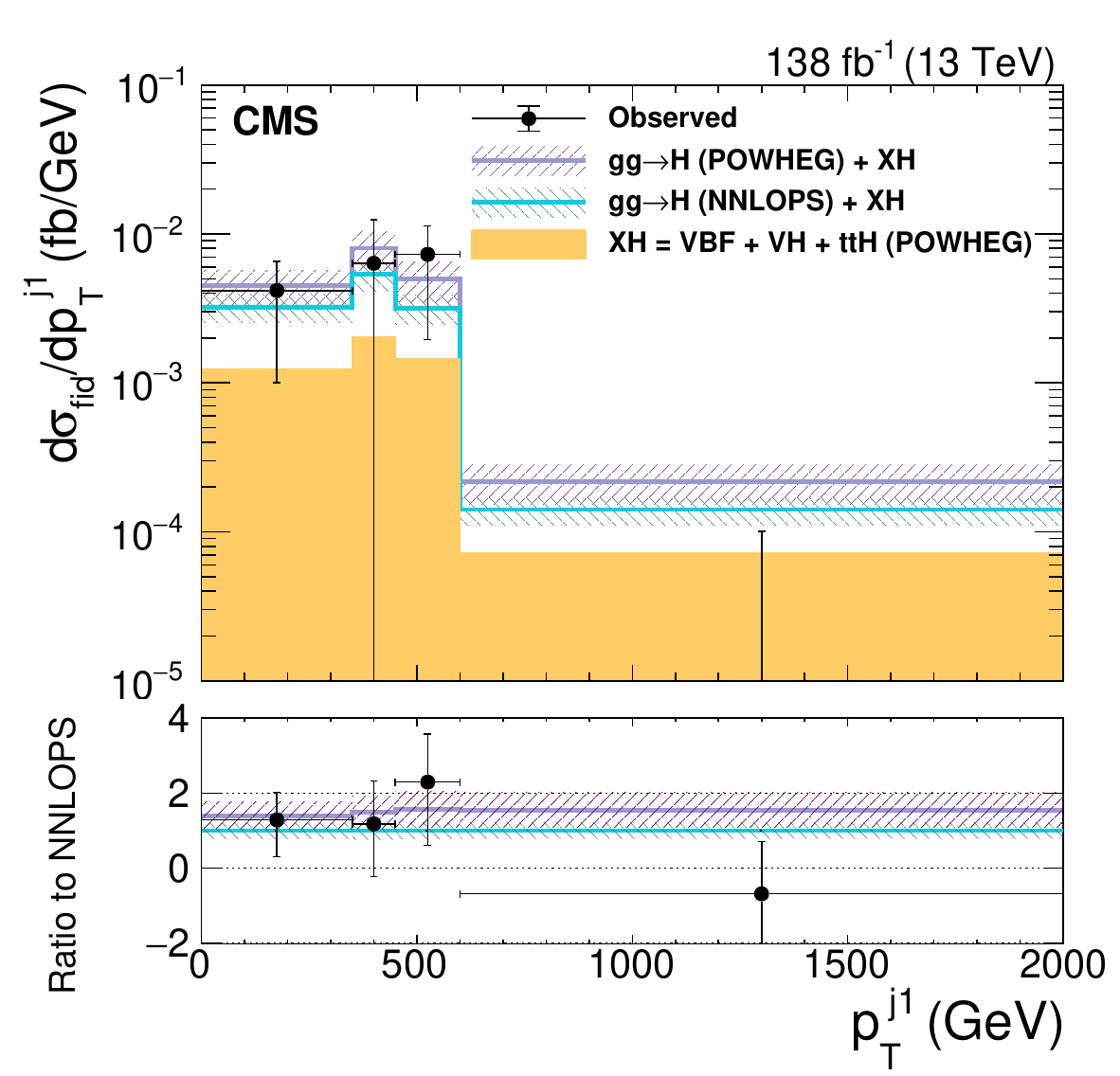}}
\end{minipage}
\hfill
%\begin{minipage}{0.32\linewidth}
%\centerline{\includegraphics[angle=-45,width=0.7\linewidth]{figexamp}}
%\end{minipage}
\caption[]{The observed and expected differential fiducial cross-sections are presented in bins of $p_{\text{T}}^{H}$ (left) and $p_{\text{T}}^{j_1}$ (right). The last bins also account for overflow.
%The theoretical uncertainty bands include contributions from several sources, such as the parton distribution function (PDF), renormalization and factorization scales, underlying event modeling, parton showering, and the branching fraction $\mathcal{B}(H \to \tau\tau)$.
}
\label{fig:Htautau_differential_boosted}
\end{figure}

In this section, STXS and fiducial differential cross section measurements \cite{CMS:2022kdi,CMS:2021gxc,CMS:2024jbe} are reported in $\mathrm{H}\rightarrow{\rm \tau}{\rm \bar{\tau}}$ final state. Studies are performed in three categories of Higgs production for which 4 channels are considered. Majority of the background is estimated from data with the use of the Tau Embedding and Fake Factor methods.  More details on the measured STXS (both by cut-based and Neural Network strategies) is reported in Ref. \cite{CMS:2022kdi}. Fiducial differential measurements have been shown in Figure. \ref{fig:Htautau_differential_resolved} and \ref{fig:Htautau_differential_boosted} for resolved and boosted regime respectively.

\begin{figure}
\begin{minipage}{0.325\linewidth}
\centerline{\includegraphics[width=0.97\linewidth]{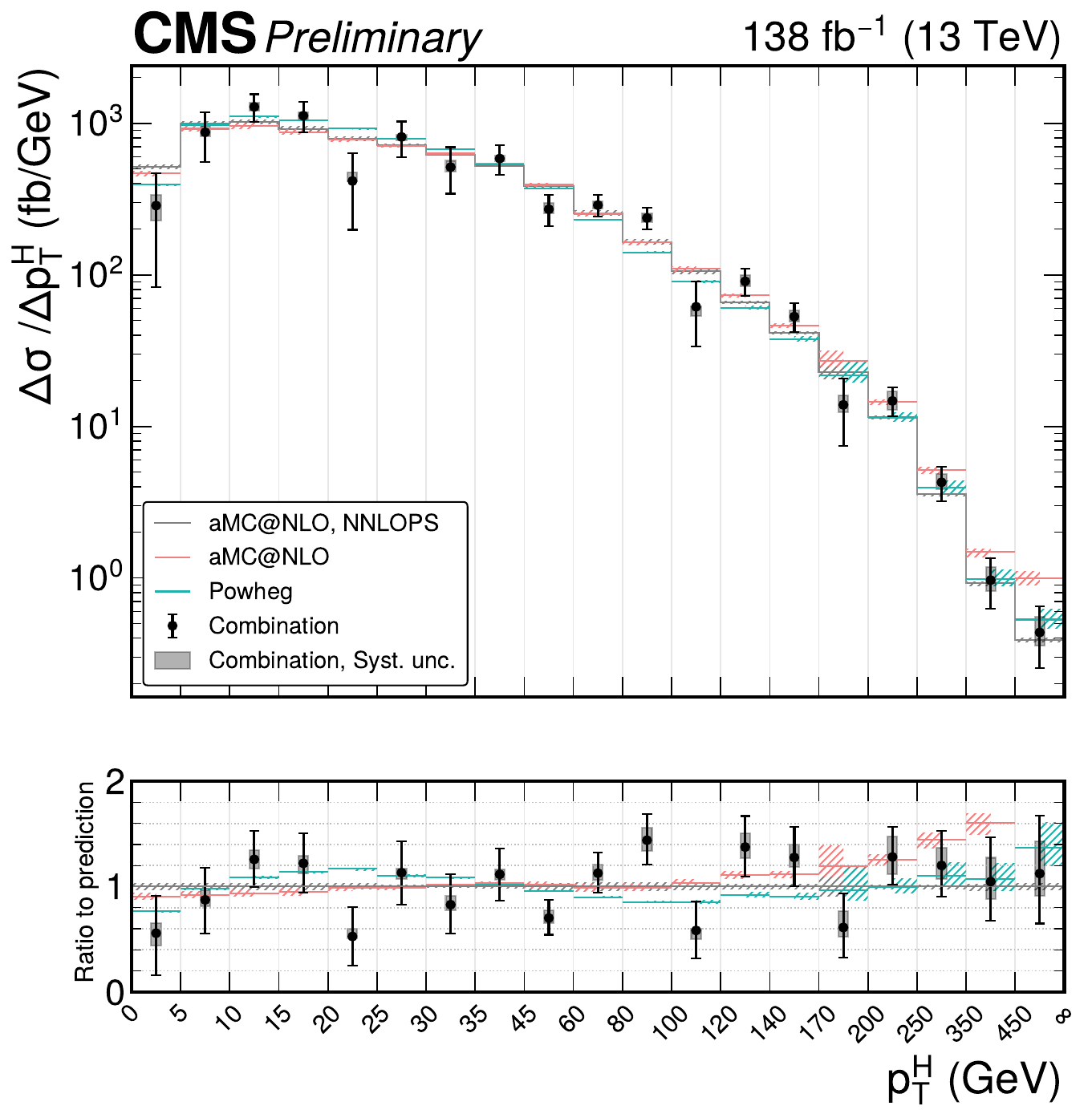}}
\end{minipage}
\hfill
\begin{minipage}{0.325\linewidth}
\centerline{\includegraphics[width=0.97\linewidth]{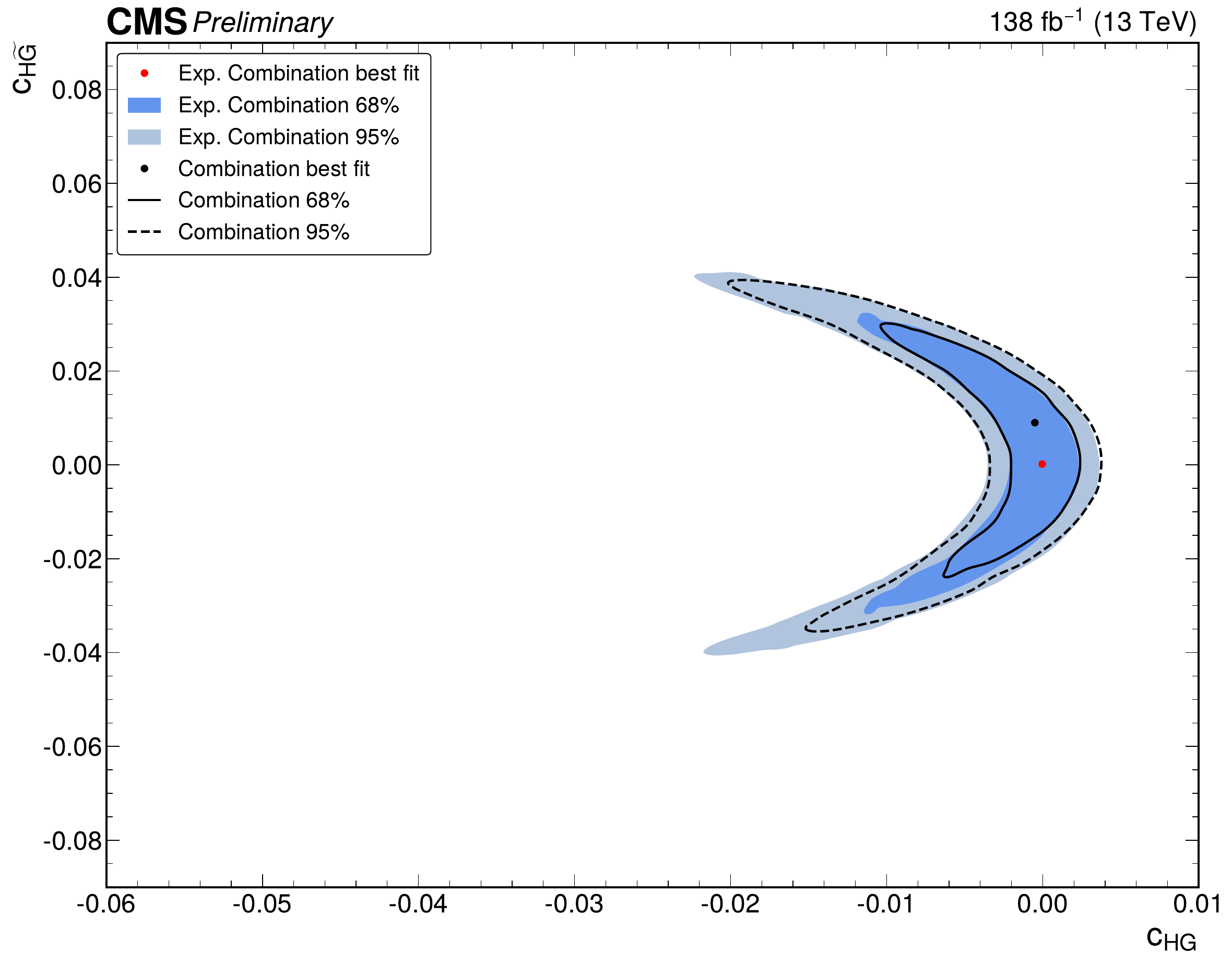}}
\end{minipage}
\hfill
\begin{minipage}{0.325\linewidth}
\centerline{\includegraphics[width=0.97\linewidth]{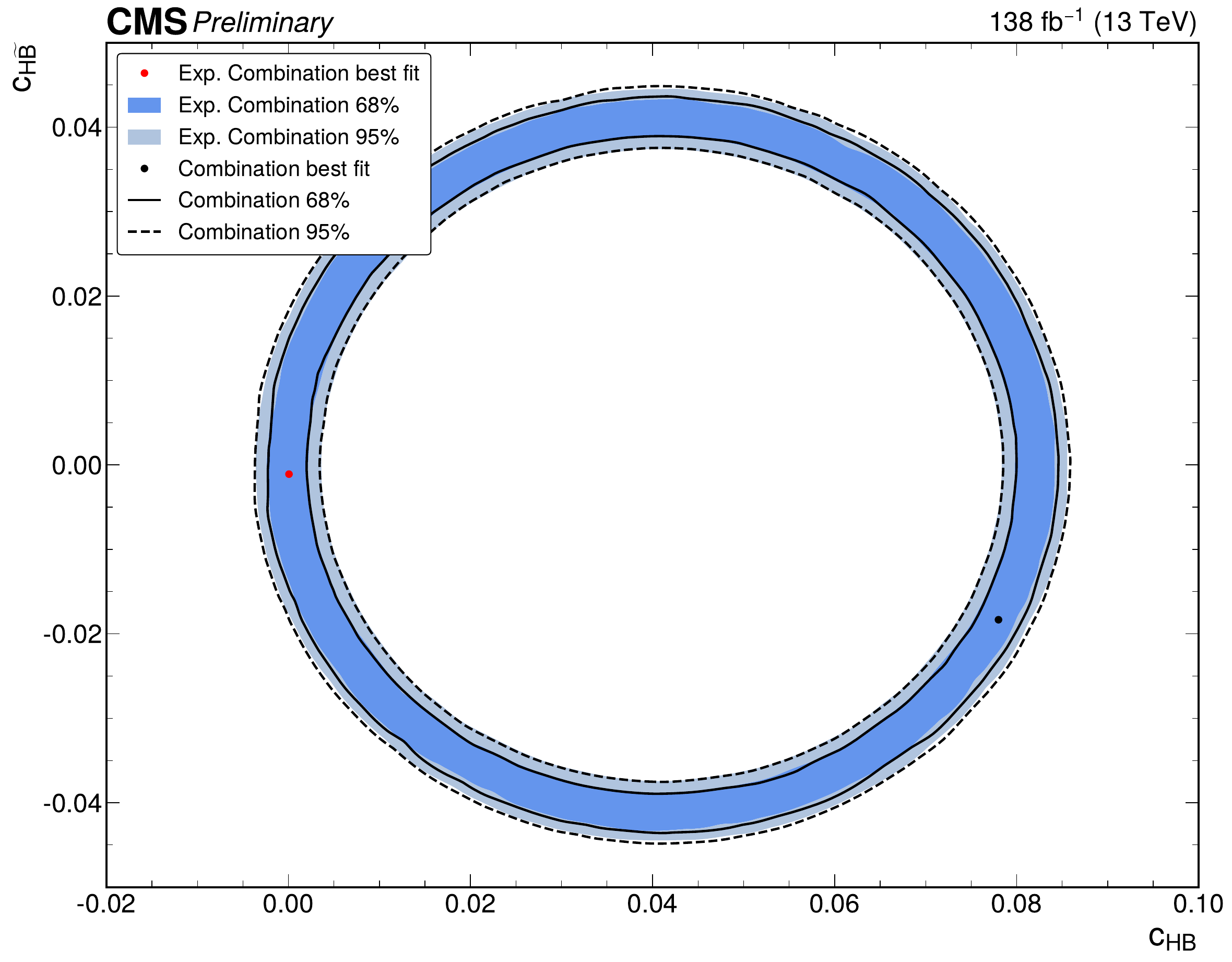}}
\end{minipage}
\caption[]{The total differential cross-section is measured as a function of $p_{\text{T}}^{H}$ (left). 
%The combined spectrum is represented by black points, with error bars corresponding to the 68\% confidence interval. The systematic uncertainty component is illustrated in gray. The Standard Model (SM) prediction is provided for different event generators. The rightmost bins of the distributions represent overflow bins and are normalized relative to the width of the second-to-last bin. In cases where the systematic uncertainty band is asymmetric and covers only one side of a data point, the systematic uncertainty on the opposite side is considered negligible. 
The observed and expected two-dimensional scans are presented for the parameter pairs $c_{HG}$–$\tilde{c}_{HG}$ (left), $c_{HB}$–$\tilde{c}_{HB}$ (right). These scans exploit $p_{\text{T}}^H$ spectra across all decay channels.}
\label{fig:combine_pTH_eft}
\end{figure}

\section{Higgs combined measurements in fermionic and bosonic final states}
In this section, combined spectra derived from measurements \cite{CMS:2024xsa} of four dominant Higgs boson decay channels ($\gamma\gamma$, $ZZ$, $WW$, and $\tau\tau$) at $\sqrt{s} = 13$ TeV, corresponding to an integrated luminosity of 138 fb$^{-1}$ is described. 
The fiducial spectra are extrapolated to the full phase space. The combined measurements are subsequently used to set constraints on Higgs couplings within the SMEFT. In cases where the bin boundaries of an observable differ across final states, they are combined using a dedicated procedure, and most of systematic uncertainty sources are kept correlated among the channels. In the Figure. \ref{fig:combine_pTH_eft}, combine differential measurement for $p_{T}^{H}$ observable is shown, which is then used to extract limits on Higgs couplings in the framework of SMEFT. %Such measurement is done by fitting CP-even CP-odd pairs of coefficients using transverse momentum of Higgs ($p_{T}^{H}$).
For the detailed results, please refer to the document \cite{CMS:2024xsa}.

\bibliographystyle{unsrt}

\section*{References}

\begin{thebibliography}{99}

%\cite{ATLAS:2012yve}
\bibitem{ATLAS:2012yve}
G.~Aad \textit{et al.} [ATLAS],
``Observation of a new particle in the search for the Standard Model Higgs boson with the ATLAS detector at the LHC,''
Phys. Lett. B \textbf{716}, 1-29 (2012)
doi:10.1016/j.physletb.2012.08.020
[arXiv:1207.7214 [hep-ex]].
%\cite{CMS:2012qbp}
\bibitem{CMS:2012qbp}
S.~Chatrchyan \textit{et al.} [CMS],
``Observation of a New Boson at a Mass of 125 GeV with the CMS Experiment at the LHC,''
Phys. Lett. B \textbf{716}, 30-61 (2012)
doi:10.1016/j.physletb.2012.08.021
[arXiv:1207.7235 [hep-ex]].
%\cite{CMS:2024pfc}
\bibitem{CMS:2024pfc}
 [CMS],
``Measurements of inclusive and differential Higgs boson production cross sections at $13.6~\mathrm{TeV}$ in the $\mathrm{H} \rightarrow \gamma\gamma$ decay channel,'' 
CMS-PAS-HIG-23-014.

\bibitem{CMS:2025wnr}
V.~Chekhovsky \textit{et al.} [CMS],
``Measurements of Higgs boson production cross section in the four-lepton final state in proton-proton collisions at $\sqrt{s}$ = 13.6 TeV,''
[arXiv:2501.14849 [hep-ex]]. 

\bibitem{CMS:2023vzh}
A.~Tumasyan \textit{et al.} [CMS],
``Measurement of simplified template cross sections of the Higgs boson produced in association with W or Z bosons in the H\textrightarrow{}bb\textasciimacron{} decay channel in proton-proton collisions at s=13\,\,TeV,''
Phys. Rev. D \textbf{109}, no.9, 092011 (2024)
doi:10.1103/PhysRevD.109.092011
[arXiv:2312.07562 [hep-ex]].
\bibitem{CMS:2024ddc}
A.~Hayrapetyan \textit{et al.} [CMS],
``Measurement of boosted Higgs bosons produced via vector boson fusion or gluon fusion in the H \textrightarrow{}$ \textrm{b}\overline{\textrm{b}} $ decay mode using LHC proton-proton collision data at $ \sqrt{s} $ = 13 TeV,''
JHEP \textbf{12}, 035 (2024)
doi:10.1007/JHEP12(2024)035
[arXiv:2407.08012 [hep-ex]].
%\cite{CMS:2023vtj}
\bibitem{CMS:2023vtj}
 [CMS],
``Measurement of the $\mathrm{t\overline{t}H}$ and $\mathrm{tH}$ production rates in the $\mathrm{H}\to\mathrm{b\overline{b}}$ decay channel with $138\,\mathrm{fb}^{-1}$ of proton-proton collision data at $\sqrt{s}=13\,\mathrm{TeV}$,''
CMS-PAS-HIG-19-011.
%\cite{CMS:2022kdi}
\bibitem{CMS:2022kdi}
A.~Tumasyan \textit{et al.} [CMS],
``Measurements of Higgs boson production in the decay channel with a pair of $\tau $ leptons in proton\textendash{}proton collisions at $\sqrt{s}=13$ TeV,''
Eur. Phys. J. C \textbf{83}, no.7, 562 (2023)
doi:10.1140/epjc/s10052-023-11452-8
[arXiv:2204.12957 [hep-ex]].
\bibitem{CMS:2021gxc}
A.~Tumasyan \textit{et al.} [CMS],
``Measurement of the inclusive and differential Higgs boson production cross sections in the decay mode to a pair of $\tau$ leptons in pp collisions at $\sqrt{s} = $ 13 TeV,''
Phys. Rev. Lett. \textbf{128}, no.8, 081805 (2022)
doi:10.1103/PhysRevLett.128.081805
[arXiv:2107.11486 [hep-ex]].
%\cite{CMS:2024jbe}
\bibitem{CMS:2024jbe}
A.~Hayrapetyan \textit{et al.} [CMS],
``Measurement of the production cross section of a Higgs boson with large transverse momentum in its decays to a pair of \ensuremath{\tau} leptons in proton-proton collisions at s=13TeV,''
Phys. Lett. B \textbf{857}, 138964 (2024)
doi:10.1016/j.physletb.2024.138964
[arXiv:2403.20201 [hep-ex]].
%\cite{CMS:2024xsa}
\bibitem{CMS:2024xsa}
 [CMS],
``Combination and interpretation of fiducial differential Higgs boson production cross sections at $\sqrt{s}=13~\mathrm{TeV}$,''
CMS-PAS-HIG-23-013.


\end{thebibliography}

\section*{Acknowledgments}
We thank School of Physics, Beihang University for providing the necessary resources and facilities under Grant 111167. We also thank the organizers of Blois2024 conference for the opportunity to present this report.

\end{document}